\documentclass{article}
\pdfoutput=1
\usepackage{balance}
\usepackage{url}
\usepackage{graphicx}
\newtheorem{proposition}{Proposition}

\usepackage{algorithm}
\usepackage{subfigure}\usepackage{algorithmic}
  \usepackage[small,compact,sf]{titlesec}

\newcommand{\komment}[1]{}
\usepackage{color}

\newcommand{\zap}{\komment}

\usepackage{amssymb}
\newenvironment{proof}{\textit{Proof}.}{$\square$}

\title{Removing Manually Generated Boilerplate from
Electronic Texts: Experiments with Project Gutenberg 
e-Books\footnote{Expanded version of \cite{kase:cascon07}}\\
{\small Department of CSAS, UNBSJ\vspace*{-1em} 
Technical Report TR-07-001
}}

\author{Owen Kaser\\University of New Brunswick
\and Daniel Lemire\\Universit\'e du Qu\'ebec \`a Montr\'eal}

\begin{document}

\maketitle

\begin{abstract}
 Collaborative work on unstructured or semi-structured documents,
such as in literature corpora or source code, often involves
agreed upon templates containing metadata. These templates 
are not consistent across users and over time. Rule-based
parsing of these templates is expensive to maintain and tends to fail
as new documents are added.
Statistical techniques based on frequent occurrences
have the potential to identify automatically
a large fraction of the templates, thus reducing the
burden on the programmers. 
We investigate the case of the Project Gutenberg\texttrademark\  corpus,
where most documents are in ASCII format with preambles and epilogues
that are often copied and pasted or manually typed. We show that
a statistical approach can solve most cases though
some documents require knowledge of English.
We also survey various technical solutions that make
our approach applicable
to  large data sets.
\end{abstract}

%

\section{Introduction}

The Web has encouraged the wide distribution of collaboratively 
edited collections of text documents. An example
is Project Gutenberg\footnote
{Project Gutenberg is a registered trademark of the Project Gutenberg
Literary Archive Foundation.
}~\cite{Gutenberg} (hereafter PG),
the oldest digital library, containing over 20,000~digitized books.
Meanwhile, automated text analysis is becoming more common.
In any corpus of unstructured text files, 
including source code~\cite{DBLP:journals/spe/AtkinsonG06}, 
we may find that some uninteresting
``boilerplate'' text coexists with interesting text that we wish to process.
This problem also exists when trying to ``scrape'' information from Web
pages~\cite{Gibson2005}.  We are particularly interested in cases where
no single template generates all text files --- rather,
there is an undetermined number and we do not initially know which template
was used for a particular file. Some templates may differ only in trivial ways,
such as in the use of white space, while other differences can be substantial
--- as is expected when distributed teams edit the files over several years.

Ikeda and Yamada~\cite{iked:text-from-templates} propose 
``substring amplification'' to cluster files according to the
templates used to generate them.  The key observations are that chunks of text
belonging to the template appear repeatedly in the set of files, and
that a suffix tree can help detect the long and frequent strings. 

Using this approach with PG is undesirable
since the suffix array would consume  much memory and require much
processing: the  total size of the 
files is  large and growing
(see Tables~\ref{gutgrowth} and \ref{sizestable}). 
Instead,
we should use our domain knowledge: 
the boilerplate in PG is naturally organized in lines and only appears at the
beginning or end of a document. 
We expect to find similar
patterns in other hand-edited boilerplate.

\begin{table}\centering
\begin{tabular}{cc}
\hline year  &  new eBooks \\ 
\hline  2001 & 1240  \\ 
\hline  2002 & 2441  \\
\hline  2003 & 4164  \\
\hline  2004 & 4049  \\
\hline   2005 & 3186  \\
\hline   2006 & 4314  \\
\hline 
\end{tabular} 
\caption{\label{gutgrowth}New PG e-Books per year~\protect\cite{gute:2007newsletter5}}
\end{table}

\subsection{Contribution and Organization}

Section~\ref{strippingPG} presents the PG corpus and describes the
difficulties in removing the preambles and epilogues, 
boilerplate that has been inserted manually and has evolved
over time.  
Given the widespread use of PG, a solution to this specific problem is
itself valuable.
We also present a few ideas that can be leveraged; for instance,
line breaking in PG boilerplate is consistent
throughout the corpus.
Section~\ref{sec:algo} presents the general frequent-lines algorithm we used.
We show analytically that our algorithm is resilient with respect
to line-classification errors in Subsection~\ref{sec:classerrors}.
In Subsection~\ref{data-structures}, various line-frequency classification
strategies
are presented including exact internal memory, exact
external memory, checksumming,  hashing, and hot-item
tracking. We show that for very skewed distribution, hashing
to few bits does not lead to frequent collisions and thus,
it can be both accurate and fast. We also provide
evidence that hot-item tracking is easier with highly
skewed distributions. Finally, in Subsection~\ref{heuristic-rules}, we present
a few simple rule-based heuristics that can help
improve our algorithm's results. We conclude in Section~\ref{exper}
with an experimental evaluation of accuracy and processing
speed. Benchmarking against an existing tool, we find 
our approach gives  superior results.

Though we have tailored the frequent-lines approach for PG, 
 it could be similarly tailored for other corpora
with manually generated or evolving boilerplate.

\subsection{Related Work}


Stripping unwanted and often repeated
content is a common task. Frequent patterns  in text documents 
have been used for plagiarism detection~\cite{1193309},
\zap{ for automatically
finding instances of copy-paste and related bugs in source code~\cite{li2004cmt},
for detecting functions that have changed name~\cite{kim2005fct},} 
for document fingerprinting~\cite{winnowing}, 
\zap{authorship attribution~\cite{hoover2001},}
for removing templates in HTML documents~\cite{1067065},
and for
spam detection~\cite{segal2004sea}.
Template detection in HTML pages has been shown to improve document 
retrieval~\cite{1141534}.

The algorithmics of finding frequent items or patterns has
received much attention. For a survey of the stream-based
algorithms, see Cormode and Muthukrishnan~\cite[p. 253]{corm:whats-hot}.
Finding frequent patterns robustly is possible using gap constraints~\cite{ji2005mmd}.

The specific problem of detecting preamble/epilogue templates in the PG
corpus has been tackled by several hand-crafted rule-based 
systems~\cite{atki:newgut-software,GutenMark,weas:weasel-reader-home}.

\section{Stripping PG}
\label{strippingPG}

\begin{table}[tb]\centering
\begin{tabular}{ll}\hline
Item & Typical size\\ \hline
Desktop memory & 1\,GiB\\ \hline
PG e-books     & 20,000\\
PG files       & 100,000\\
PG .txt files  & 25,000\\
PG  size       & 150\,GiB\\
PG .txt size   & 10\,GiB\\
PG .txt size   & 200\,M lines\\
PG line length & 50 characters, average\\
PG avg file size & 8500 lines\\ \hline
First DVD (2003): \\\hline
 ebooks  & 10,000\\
 .txt files & 11,000\\ 
 .txt size & 5\,GiB\\ 
 .txt size & 80\,M pre-proc.\ lines\\ 
 .txt size   & 55\,M distinct pre-proc.\ lines\\
 .txt size   & 10,000 freq.\ pre-proc.\ lines\\
\hline
\end{tabular}             
\caption{\label{sizestable}Typical size assumptions for 2007.
Preprocessing, discussed in \S~\ref{algo-details}, removes
short lines.}
\end{table}

The PG corpus contains texts in different formats, and
subsets of the corpus are distributed on various media.
An e-book may be
available in several formats and may have been revised (for instance,
to correct errors) several times.  Furthermore, a given book may exist in more
than one edition, since published books are frequently revised by their authors
over time. 
An \emph{e-book file} corresponds to a certain format, edition and version
of an e-book, and this  paper is  primarily concerned with e-book files, 
rather than e-books themselves.  The number of e-book files in PG
and its first DVD are given in Table~\ref{sizestable}, along with
related statistics.
While some of these sizes were given by PG itself~\cite{GutenbergMirror}, 
others were obtained by processing the DVDs,  whereas yet
other were extrapolated from a random sample of approximately
100~e-books.

Due to the size of the full corpus, 
we report on results from the first and second
PG DVDs~\cite{GutenbergDVDs}.
These DVDs contain most of the current text files and can be viewed as 
snapshots of the text portion of the archive in 2003 and 2006.
With few exceptions, e-books produced by PG are available in
unstructured plain text, whereas only a smaller subset are available as XML or HTML.
Therefore, projects that wish to use data from PG need to work with 
unstructured text.  However, the method by  which PG e-books are created
produces some challenges.

\begin{figure}
\centering
\includegraphics[width=.6\textwidth]{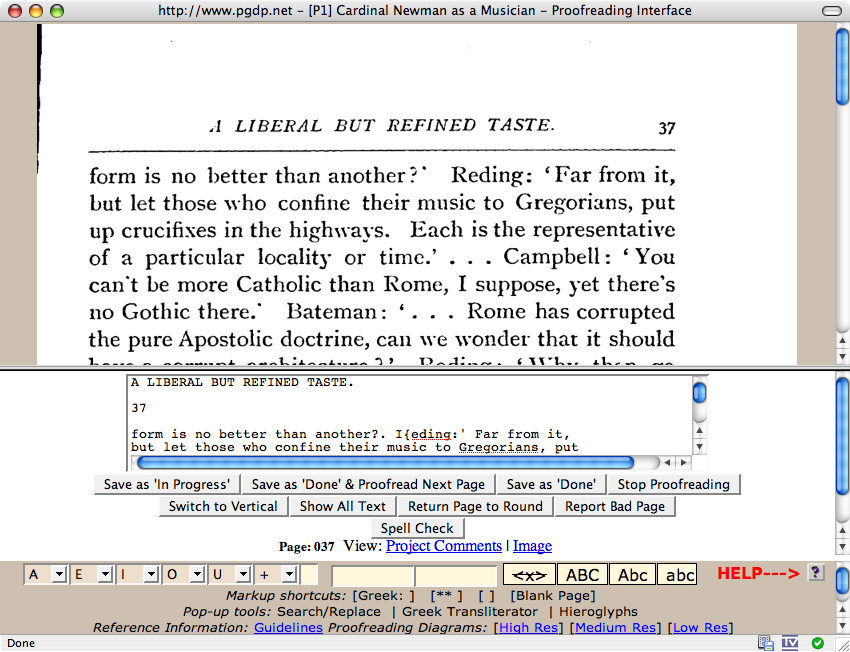} 
\caption{\label{fig:dp}Screen shot of the ``Distributed Proofreaders'' application. }
\end{figure}

Although the first PG e-books were undoubtedly
typed in manually, the production of e-books has been modernized although
remaining manual. The original
documents are typically in a printed form and have
been scanned. Volunteers then
use OCR  and proof-reading tools such as the
``Distributed Proofreaders''~\cite{dp2003} 
application (see Fig.~\ref{fig:dp}).
The volunteers merely translate a printed document into an electronic text
document: headers, tables, sections, foreign text, and so on are not marked up
in the plain-text version.
  At some stage, boilerplate text is added.

With PG e-books, there is always a
preamble that provides various standard metadata, possibly
comments from the transcribers,
comments about the legal status of the text, and so forth.  Following the
transcribed body of the book, there is frequently an epilogue that 
contains further information about PG, the transcribers, 
and so forth.  Text-analysis tools are typically applied only to the body
of the book; clearly, a vocabulary analysis will not wish to report on
vocabulary that comes from the preamble. 
We want an automated solution to remove the
preamble and epilogue, and we are willing to accept less than perfect
accuracy in return for fast runtimes.

One additional complication is that the preamble and epilogue 
portions are not entirely one
block of boilerplate.
The names of the transcribers, the title of the book, and other variable items
appear in the preamble and epilogue.  Further,  it appears that the preamble
and epilogues are typically the result of human-guided 
cut-and-paste (or even re-typing) activities.

PG does specify how the preamble and the epilogue should be
formatted. An e-book file \textit{should} always begin with 
a specified template~\cite{GutenbergHeaderFooter}:


\begin{small}
\begin{verbatim}
The PG eBook, (insert title), by (insert author)

This eBook is for the use of anyone anywhere at no cost and with
almost no restrictions whatsoever.  You may copy it, give it away or
re-use it under the terms of the PG License included
with this eBook or online at www.gutenberg.org

** This is a COPYRIGHTED PG eBook, Details Below **
**     Please follow the copyright guidelines in this file.     **


Title: (insert title)

Author: (insert author)

Release Date: (insert date)  [eBook #(insert ID number)]

Language: (insert language)

Character set encoding: (insert character encoding)


*** START OF THE PROJECT GUTENBERG EBOOK (insert title)***



   ----------------------------------------
   ---- the body of the text goes here ----
   ----------------------------------------



*** END OF THE PROJECT GUTENBERG EBOOK (insert title)***

******* This file should be named (insert file name) *******

(...)
\end{verbatim}\end{small}
However, few files satisfy this specification; for instance,
none of the first 10 files posted on February 27, 2007 matched.  
In most cases, the first line lacked a comma and an occurrence of
`THE' had been replaced by `THIS'.
Further, thousands of e-book files mention some of the
old Internet hosts for PG, and these were not updated when
the host changed:  only two e-book files in the first 
DVD refer to \texttt{www.gutenberg.org}, 
and then not in the position required by the specification.  Similarly, on
the second DVD at least 13,000 (and probably more) of the
roughly 20,000 files on the second DVD also do not match.  

Besides the problem that the specification refers to items (such as
Internet hosts) that change over time, the specification is not meant 
for automated processing. 
And the specification is not even complete:
can we separate authors by new lines or just comas or 
other punctuation?   How are editors handled?

The desired preambles and epilogues used in PG e-book files have changed
several times over the years, and they may change again in future.
This makes fully hand-crafted PG
parsers~\cite{atki:newgut-software,GutenMark,weas:weasel-reader-home}
an unsatisfactory solution.  We feel the best way to obtain a robust
solution is to use  methods that
automatically adjust to changes in data.

\begin{table}\centering
\begin{tabular}{ccc}
\hline  $n$ & number of distinct $n$-grams & error (\%) \\ 
\hline 10 & 1.06 $\times 10^9$ & 14   \\ 
\zap{  9 & 1.09 $\times 10^9$   & 14  \\ }
  8 & 1.01 $\times 10^9$   & 14    \\ 
\zap{ 7& 1.05$\times 10^9$  & 14 \\ }
  6& 1.06 $\times 10^9$ & 14   \\ 
  5& 8.85$\times 10^8$ & 14   \\ 
  4&  7.99$\times 10^8$ & 16   \\ 
 3& 4.97 $\times 10^8$  & 20  \\ 
 2& 1.62 $\times 10^8$ & 37 \\ 
 1 & 1.8 $\times 10^7$ & 100\\
\hline
\end{tabular} 
\caption{\label{ngramcounts} Estimated number of distinct $n$-grams in the texts of the second PG DVD.}
\end{table}

\subsection{Domain Knowledge}

Several properties of PG can help us choose an approach
detect boilerplate.  First,
we know that all PG e-book files have preambles, although not all
have epilogues.   
Second, we observe that
the boilerplate text-flow is typically the same throughout the
documents:  sentences are split
into lines at the same locations. This means that we can process frequent text-file lines,
rather than parsing sentences from the text and looking for frequent
sentences.  We found no solution improvement and a notable
running-time increase when we tried looking for frequent sentences. 
Third,
we can set reasonable bounds on the
maximum size of preamble and epilogue.
Fourth, we can estimate the maximum number of times that a non-trivial
line from the body 
of an e-book may be replicated.

We know there some textual replication between PG 
e-book files --- but how much?
Some e-books are compilations of an author's work, whereas
some of the items in the compilation are sometimes themselves e-books.
Considering the two PG DVDs,  there appears to be more replication in 
first PG DVD, where
we find over 700 e-books present in both the initial
and a second 
(corrected) version.
Eighty e-books appear in a third version. 
\textit{Frankenstein} is an extreme case: the first DVD contains
two editions of the book:  one has five versions on the
DVD whereas the other has three, for a total
of 8~similar e-books.  
We estimate that nearly 10\% of the first DVD's e-book files are near-duplicates.  However, we do not
expect to see more than 10 files that are (near) copies, for any given e-book.

Finally, the properties of PG discourage us from considering frequent
$n$-gram approaches, where an $n$-gram (of words) is a consecutive sequence
of $n$ words.  To detect transitions in documents,
$n$-gram statistics are sometimes
used~\cite{1193309}. For example, the 5-gram ``END OF THE PROJECT
GUTENBERG'' is likely to be frequent in the PG corpus. Computing the
number of occurrences of each $n$-gram can be done efficiently with
suffix arrays using the Nagao-Mori algorithm~\cite{NagaoMori}.  The
size of the suffix array, which is several times larger than the
original corpus, seems an unnecessary burden when one's purpose is to
remove boilerplate.  Alternative data structures that maintain counts for
each distinct $n$-gram  are unlikely to be much better
as we determined~\cite{viewsizetechreport} there are nearly $10^9$ distinct $5$-grams on
the second PG DVD by defining a word as text separated by white space (see Table~\ref{ngramcounts}).

\section{Algorithm}
\label{sec:algo}

Our solution
 identifies frequent lines of text in the first
and last sections of each file. These frequent lines are recorded
in a common data structure.
Then, each file is processed
and the prevalence of infrequent lines is used to detect a transition
from a preamble to the main text, and one from the main text to an
epilogue.

\begin{figure}[htb]\centering
\includegraphics[width=0.6\columnwidth]{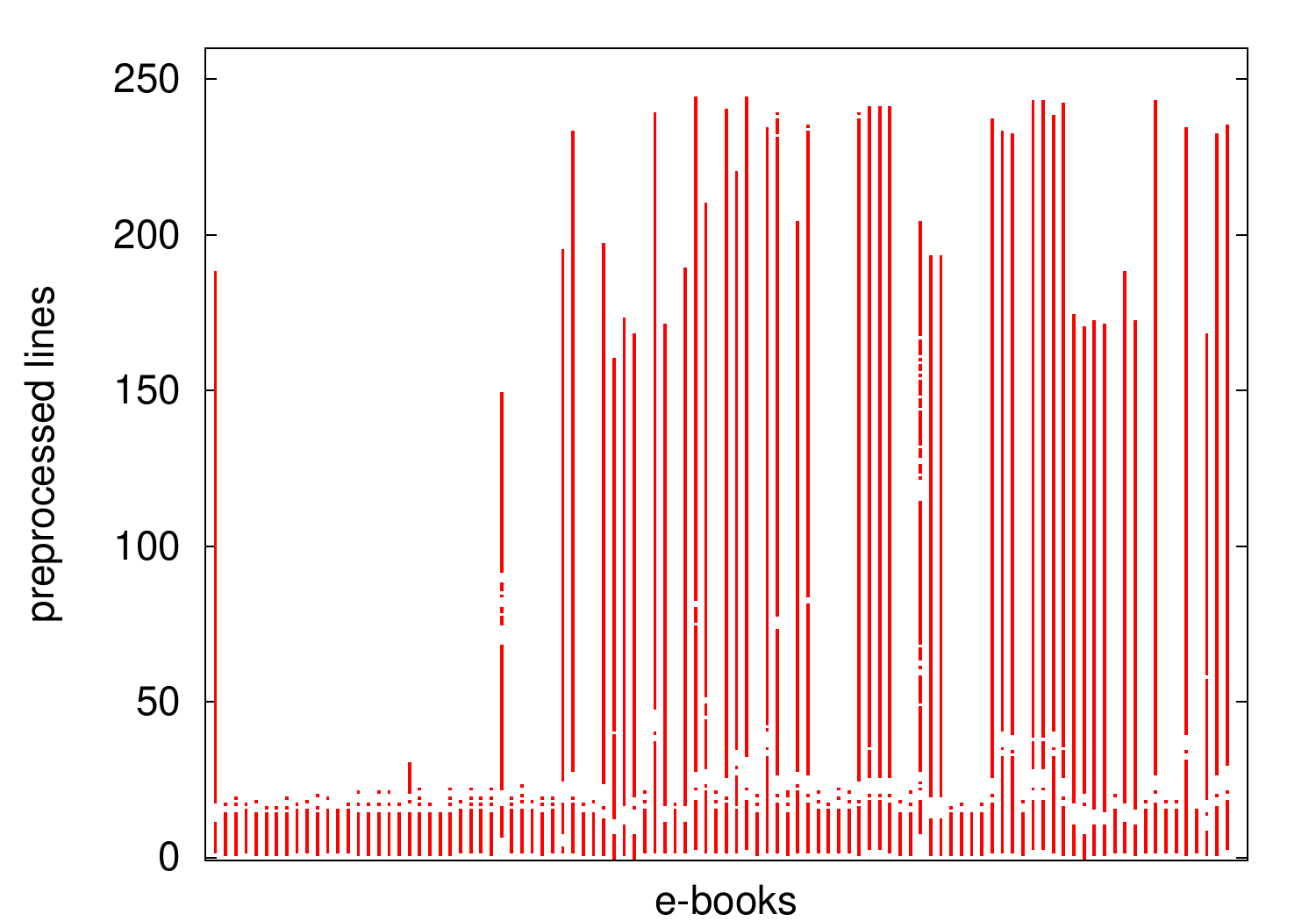}
\caption{\label{boiler-vis} Pre-processed lines of frequency
100+ in the first DVD. The $y$-axis measures (pre-processed) line positions
within an e-book file.  There were 100~e-book files sampled from the
first DVD.}
\end{figure}

To motivate this approach, see Fig.~\ref{boiler-vis}. It
shows the frequencies of the first 300~lines in each of
100~e-books randomly sampled from the first DVD.
From it, we see files  with long preambles (an older style) as well
as those with short preambles (used in recent e-books).

\subsection{Details}
\label{algo-details}

The core algorithm is shown in Algorithm~\ref{boiler-algo}.
A line is frequent if it occurs more than $K$~times 
in the corpus: we chose $K=10$.
We used the following ideas.
\begin{enumerate}\setlength{\itemsep}{0ex}\setlength{\parskip}{0ex}
\item Preamble and epilogue lines are frequent. Exceptions include
lines meant to contain the title of the work.
\item Lines not in the preamble or epilogue are infrequent with high 
probability, except
for trivial lines such as ``CHAPTER 1''.
\item Within most preambles or epilogues, there may be small gaps of
 infrequent text, but no large gaps.
\item The sizes of the preamble and epilogue can be bounded by $p_{\max}$ and 
$e_{\max}$ lines, respectively. (We use $p_{\max} = e_{\max} = 300$,  after 
removing trivial lines.)
\end{enumerate}

The algorithm always does some minor preprocessing of lines.  It trims white
space from the beginning and end of each line, and it replaces
runs of `*',  `-' or whitespace characters by `***', `{-}{-}{-}' or ` ',
respectively.  It then omits as trivial those lines that are
short (less than 30~characters) or have no alphabetic characters.  These rules reflect
the variety of typing discrepancies that seem likely when
a human typist attempts manual duplication of boilerplate, or
from ``boundary errors'' during cut-and-paste sessions.
Omitting short lines should avoid problems from lines such as 
``CHAPTER ONE.''

\begin{algorithm}[htb!]
\begin{small}
\begin{algorithmic}
\STATE\textbf{Pass 1}: identify frequent versus infrequent text
\STATE $p_{\max}\leftarrow 300$,  $ e_{\max} \leftarrow 300$
\STATE Scan first $p_\textrm{max}$ and last $e_\textrm{max}$ lines of each file.
\FOR {each line seen}
\STATE  pre-process line
\STATE  record in a frequent-lines data structure
\ENDFOR
\vspace*{0.1cm}
\hrule
\vspace*{0.1cm}
\STATE \textbf{Pass 2}: identify preambles and epilogues

\FOR { each file}

\STATE  constant GAP\_MAX $\leftarrow 10$

\STATE  pre-process lines, skip till frequent line seen

\STATE  gap $\leftarrow 0$
  \WHILE { gap $<$ GAP\_MAX}
      \STATE read and pre-process line 
     \IF { line is frequent}
     \STATE  gap $\leftarrow 0$
     \ELSE 
      \STATE increment gap
      \ENDIF
\ENDWHILE
  \STATE record that the preamble ends at the last-seen frequent line

  \STATE process the lines in the file in backwards order
  \STATE  gap $\leftarrow 0$
  \WHILE {  gap $<$ GAP\_MAX}
     \STATE read previous and pre-process line in file
     \IF { line is frequent}
     \STATE  gap $\leftarrow 0$
     \ELSE 
      \STATE increment gap
      \ENDIF
\ENDWHILE
  \STATE record that the epilogue starts at the most-recently-seen
     frequent line, or that no epilogue was seen
     if no frequent line was found
\ENDFOR
\end{algorithmic}\end{small}
\caption{\label{boiler-algo}
Algorithm to detect the end and start of boilerplate in a collection of documents.
Lines that are too short or have no alphabetical characters are ignored.
}
\end{algorithm}

\subsection{Classification-Error Effects}
\label{sec:classerrors}
Since Algorithm~\ref{boiler-algo} cannot directly infer that a line
belongs to the preamble or epilogue, it approximates this by
distinguishing between frequent and infrequent lines.
This test can yield a
\emph{false positive} when a non-boilerplate line occurs frequently,
or a \emph{false negative} when a boilerplate line occurs
infrequently.  The situation becomes worse if
we choose inexact methods for determining line frequency.

Two types of errors may occur when trying to identify the 
preamble from line frequencies. If a sequence of false negatives
occurs within the preamble, then we may
cut the preamble short. If some false positives occur
shortly after the preamble, then we may overestimate
the size of the preamble. The analysis of the epilogue follows
a similar pattern.

The effect of false negatives on the underestimation 
of the preamble is difficult to assess analytically,
since 
occurrences are not independent of one another. For instance,
an infrequent line about some unusual copyright situation is
probably followed by  another infrequent line
that elaborates.  Nevertheless, in the simplistic analytic model
where false negatives occur with probability $\sigma$, the probability
of encountering GAP\_MAX false negatives is small and bounded
by $L \sigma^\textrm{GAP\_MAX}\leq 300 \sigma^\textrm{GAP\_MAX}$ where
 $L$ is the length
of the preamble in lines (known to be almost always smaller than 300).
 Even if $\sigma$ is moderately
large (25\%), the probability $300 \sigma^\textrm{GAP\_MAX}$ will be
very close to 0 for $\textrm{GAP\_MAX}\approx 10$ 
(for example, $300 \times 0.25^{10}\approx 0.0003$). In fact,
the expected number of lines before GAP\_MAX false negatives
are encountered is $\sum_{k=1}^{\mathrm{GAP\_MAX}} \sigma^{-k}$ (see 
Proposition~\ref{danprop}) which is 1.4~million lines for $\textrm{GAP\_MAX}= 10$
and  $\sigma=0.25$. 
Therefore, we might
conclude that it is unlikely we will cut the preamble short by more
than GAP\_MAX lines.  This is assessed experimentally in \S~\ref{exper}.

\begin{proposition}\label{danprop}
 If the probability of flipping a head is $p$, then 
the expected time before a consecutive sequence
of $n$ heads are observed is $\sum_{i=1}^n 1/p^i$.
\end{proposition}
\begin{proof}
 Let $H(k,n)$ be the expected time before we observe a 
sequence of $n$ heads given that we just observed a sequence
of $k$ heads. We have that
$H(k,n)=(1-p)(1+H(0,n))+p(1+H(k+1,n))=1+(1-p)H(0,n)+pH(k+1,n)$.
Starting with $H(0,n)=p(1+H(1,n))+(1-p)(H(0,n)+1)=1+(1-p)H(0,n)+pH(1,n)$ and
recursively substituting the former equation, we
get $H(0,n)=(1-p^k)H(0,n)+1+p+\ldots+p^{k-1}+p^k H(k,n)$,
which we can prove by induction. Thus, because $H(n,n)=0$, we have
the desired result: $H(0,n)=\frac{1+p+\ldots+p^{n-1}}{p^n}$.
\end{proof}

However, regarding the risk of overestimation,
it seems reasonable to assume the occurrences of false
positives occur independently with probability $p$ and an analytical
approach is more conclusive. Indeed, in the body, one line's
being frequent (such as ``CONTENTS OF THE SECOND VOLUME.'')
would rarely imply that neighbouring lines are frequent.
For simplicity, we assume that there
are no false negatives  --- false negatives 
 only help the situation
 --- and we focus on preambles.  Epilogues would be
similar. 
Let $p$ denote the probability of a false positive.
We see that
the last line of the identified preamble will be GAP\_MAX (pre-processed) 
lines \textbf{before} the point when a consecutive
sequence of GAP\_MAX negatives has been seen.  
Using Proposition~\ref{danprop},  we determine that the expected
number of misclassified lines following the preamble is
$\sum_{k=1}^{\mathrm{GAP\_MAX}} (1-p)^{-k} - \mathrm{GAP\_MAX}.$ 

\begin{figure}[htb]
\centering\includegraphics[width=0.6\columnwidth]{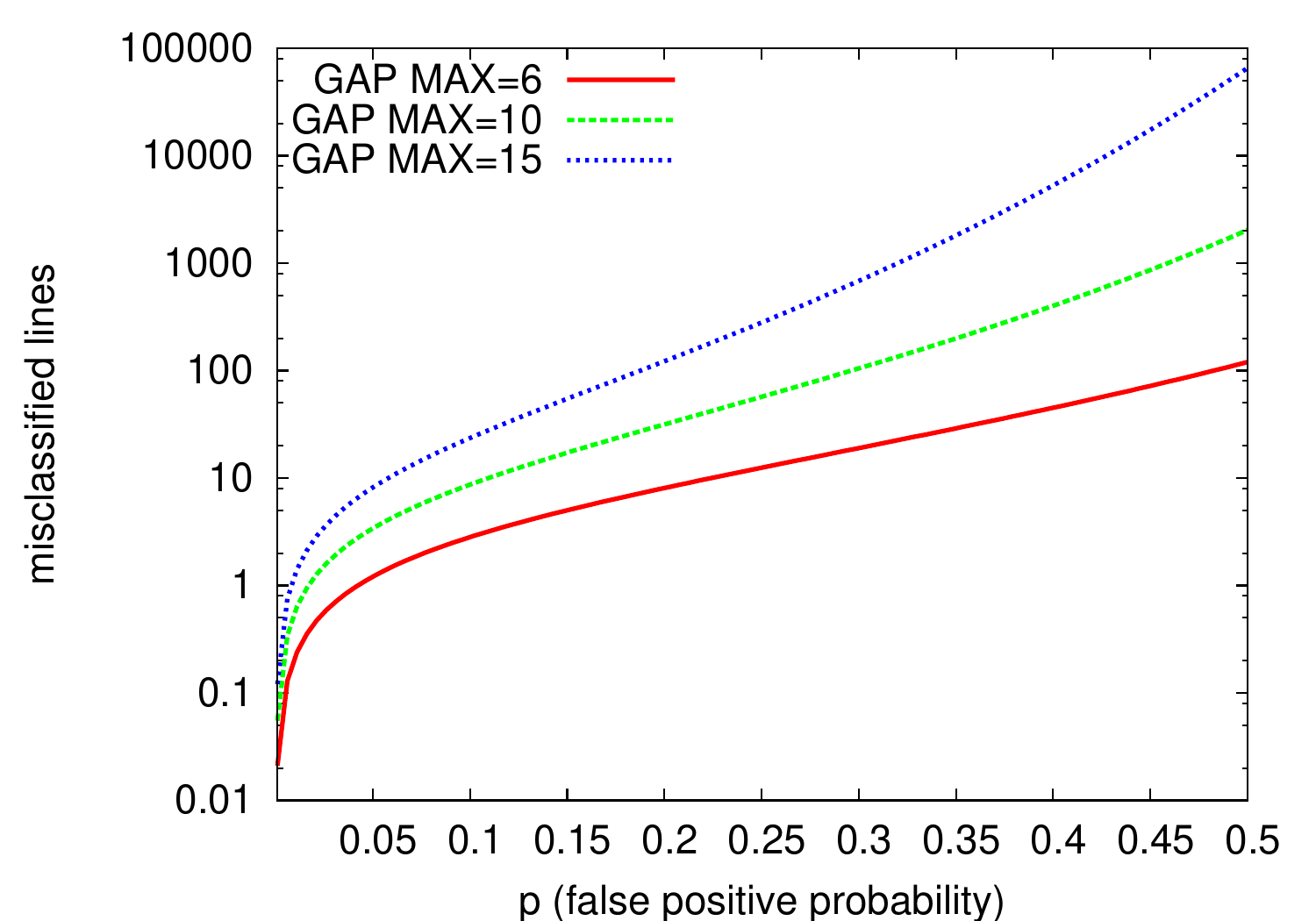}
\caption{\label{overshot-fig} Expected number of non-preamble
lines misclassified, given false-positive probability $p$.
We assume false positives are independent and show results on
a logarithmic $y$ axis.}
\end{figure}

Figure~\ref{overshot-fig} shows the effect of various values of $p$
on this function.  With MAX\_GAP = 10, we want $p \leq 20\%$. 

\subsection{Data Structures}
\label{data-structures}

The algorithm's first pass builds a data structure to identify the
frequent lines in the corpus.  Several data structures are possible,
depending whether we require exact results and how much memory we
can use.   One approach that we do \emph{not} consider in
detail is taking a random sample of the data.
If the frequent-item threshold is low (say $K=5$), too small a 
sample will lead to many new false negatives. 
However, when $K$ is large, sampling 
might be  used with any of the techniques below.

Although we assume that only 600 ($p_{\max} + e_{\max}$)  lines
are processed per PG e-book file, there may be similar applications
where this assumption cannot be made and the entire file must
be processed. The impact of removing the assumption on the desired
data structure should be considered.

\subsubsection{Exact Counts Using Internal Memory}

For exact results, we could build a hash table that maps each line
seen to an occurrence counter.  In Java, the Collections library
provides \verb+HashMap<String,Integer>+.  This has the disadvantage of
consuming much memory.  Of the 600 pre-processed lines we take from
each file, we estimate about half are 
unique, non-preamble/epilogue lines, because a given file 
normally has a large
preamble or a large epilogue, but not both.
Within the first and last 300 pre-processed 
lines of the first DVD's files, we
see only 3.4~million distinct lines.
A routine estimation (details omitted) says 
we need about 700\,MiB for our data structure.
If the data structure indeed fits, it will tie up most of
our main memory (see Table~\ref{sizestable}).

\subsubsection{Exact Counts Using External Memory} 

\begin{figure}[htb]\centering
\subfigure[\label{histo}Number of lines appearing a certain number 
of times. The scales are logarithmic.]{ 
\centering\includegraphics[width=0.6\columnwidth]{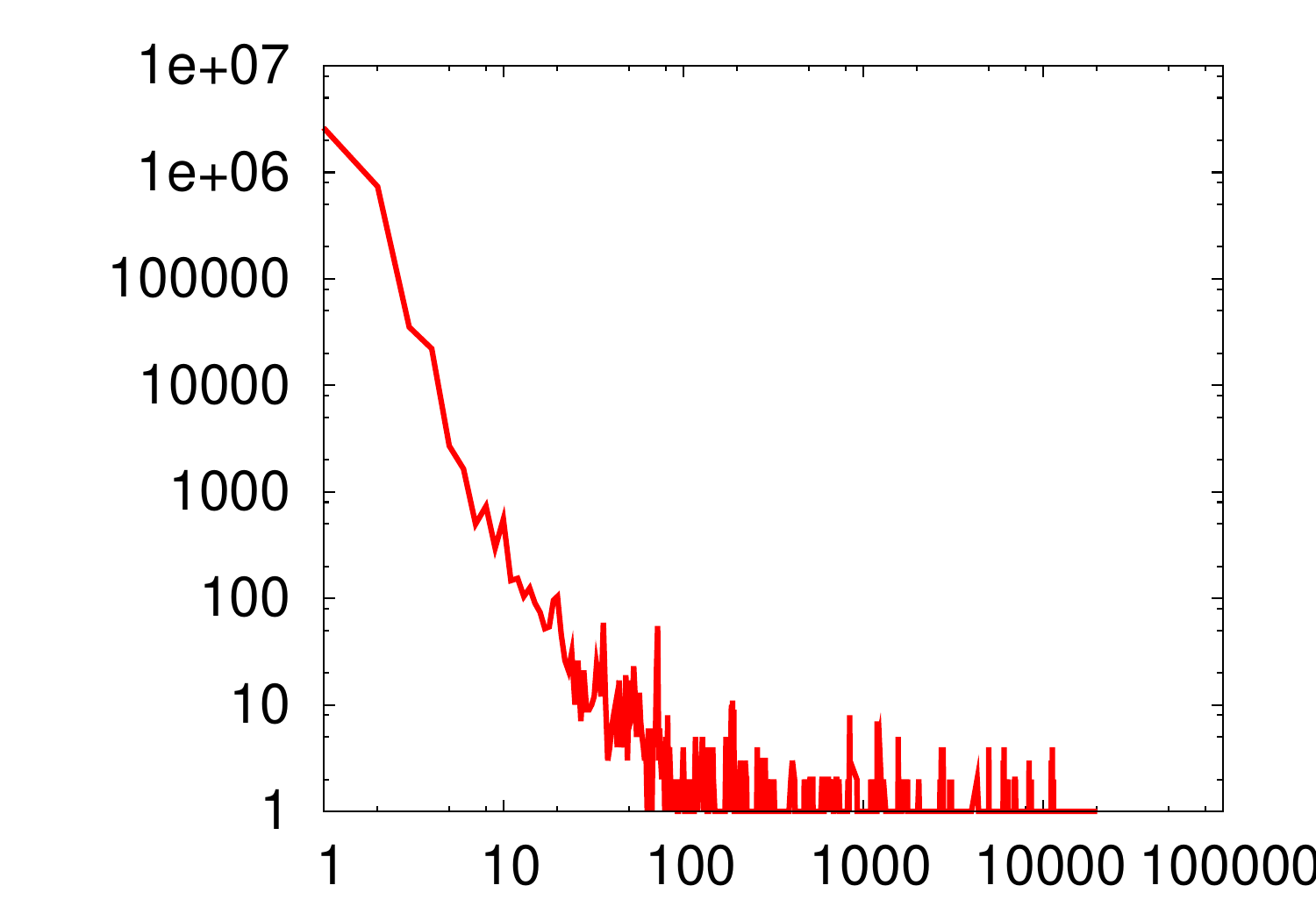}
}\\
\subfigure[\label{probahisto}Probability to pick a line
having at least the given number of occurrences.]{
\centering\includegraphics[width=0.6\columnwidth]{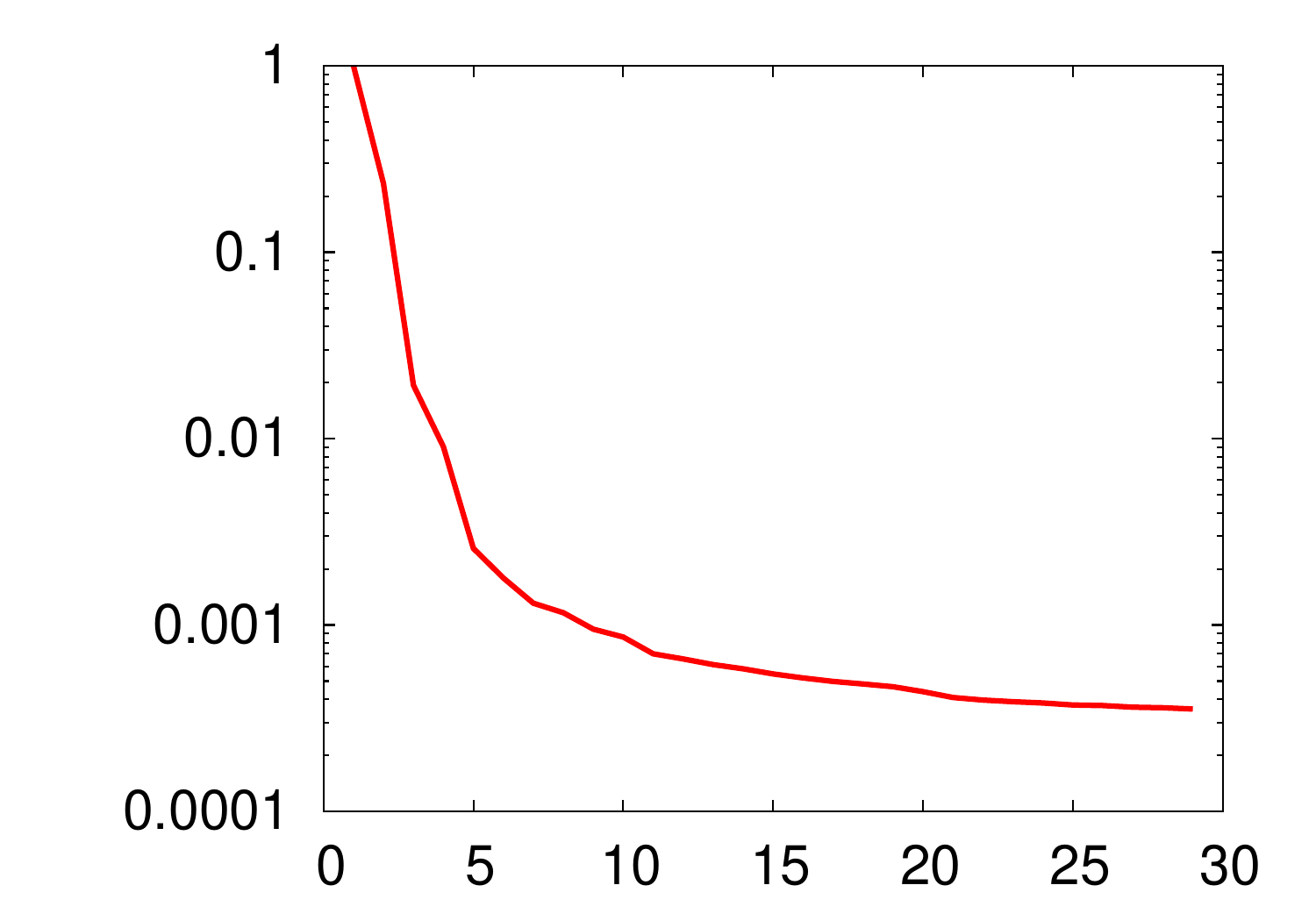}
}
\caption{Line occurrence statistics in the beginning and end 
(300 non-trivial lines)
of text documents  on the second Gutenberg DVD.
}
\end{figure}

To know exactly which lines occur
frequently, if we have inadequate 
main memory, an  external-memory solution is to sort the lines.
Then a
pass over the sorted data can record the frequent lines,
presumably in main memory.
If we  build a file \texttt{F} containing 
just the
first and last 300 non-trivial pre-processed lines of each file,  
the following GNU/Linux pipeline prints 
a list of under 3,000 frequent lines (occurring 10 times or more) in less
than 100\,s on our somewhat old server:

\begin{footnotesize}
\begin{verbatim}
sort F -S 512M | uniq -c | egrep -v '^ *[1-9] '
\end{verbatim}\end{footnotesize}
Part of the result (frequency counts and the lines) is shown in
Fig.~\ref{sample-sorted}. A histogram for 
the data (see Fig.~\ref{histo}) shows
that the distribution is very skewed: out of 3.4~million distinct lines,  
2.6~million lines appear only once and  0.7~million lines
lines appear twice, with less than 100,000 lines  appearing more than 2 times
and fewer than 3,000 appearing 10 or more times.
We see e-book-specific text occurring with 
some surprising frequency due to cut-and-paste errors (we verified that
19 files that were \emph{not} \texttt{1rbnh10.txt}, mistakenly claimed
that this should be their name).
We also see  boilerplate shifts: (\textit{eBook} and
\textit{EBook} versus \textit{etext},  \textit{net} versus \textit{gross} 
and \textit{Project} versus \textit{Foundation}).   
Only 53 files have the
\textit{twenty percent} spelled out (see Fig.~\ref{sample-sorted}), illustrating that
rare boilerplate \emph{does} occur and showing how false positives/negatives 
may arise.

\begin{figure*}
\begin{footnotesize}
\begin{verbatim}
     68 ***The Project Gutenberg's Etext of Shakespeare's First Folio***
   1034 ***These EBooks Were Prepared By Thousands of Volunteers***
   1415 ***These Etexts Are Prepared By Thousands of Volunteers!***
    126 ***These Etexts Were Prepared By Thousands of Volunteers!***
   5058 ***These eBooks Were Prepared By Thousands of Volunteers!***
     20 ***This file should be named 1rbnh10.txt or 1rbnh10.zip***
    128 (2) Pay a royalty to the Foundation of 20% of the gross
     54 (2) Pay a royalty to the Project of 20% of the net
     53 [3] Pay a trademark license fee of 20% (twenty percent) of the
   8061 [3] Pay a trademark license fee to the Foundation of 20% of the
\end{verbatim}
\end{footnotesize}
\caption{\label{sample-sorted}Some of the approximately 3,000 frequent sorted lines in the first DVD.}
\end{figure*}

\subsubsection{Checksumming}

For nearly exact results, we can hash lines to large
integers, assuming that commonly used hashing algorithms
are unlikely to generate many collisions.
We chose the standard CRC-64 checksum, and a routine 
calculation~\cite{wiki:birthday-paradox}
shows that with a 64-bit hash, we can expect to 
hash roughly $2^{64/2}$ distinct lines before getting
a collision.
With hashing, each line is represented using 8~bytes,
reducing memory usage tenfold.
Even considering the other overheads of the data structure, Java's
\verb+HashMap<Long,Integer>+,  it
seems clear that the tops and bottoms of all
PG files can be processed comfortably 
within our 1\,GiB limit.

If we hold constant the threshold $K$ beyond which an item is
considered frequent, checksumming does not increase the rate of 
false negatives. 
It may increase the rate of false positives, when
the sum of several colliding lines' frequencies exceeds $K$.
This effect should be negligible for CRC-64.

Solutions that use even less memory may be desirable, when
all of a file must be processed.  For instance,
suppose that a block of boilerplate could appear anywhere in
a file. 

\subsubsection{Hashing to Millions of Counters}
\label{billionctrs}
To use even less memory than CRC-64 hashing, one solution is to 
use a smaller hash (e.g., a 30-bit hash) and accept some collisions.
Once the range of the hash function is small enough, it can directly
index into an array of counters, rather than requiring a lookup in a
secondary structure mapping checksums to counters. 
This could
lead to speed or space advantages: for the entire
PG corpus and with $K=10$, we could use $2^{30}$ 8-bit counters for a total of 1\,GiB.
With larger $K$ ($K>255$),
approximate counting~\cite{morr:approx-counting}
can
trade count accuracy for the ability to count to $2^{2^8}$.

When using this data structure with Algorithm~\ref{boiler-algo}, \zap{upon
reading a line in the first pass, we compute 
a 30-bit hash and then increment the 
counter indexed by the hash.  
In the second pass, to determine whether a line is frequent we simply
compute its 30-bit hash and examine the indexed counter.  All} 
frequent lines will be correctly recognized as such, but we can expect
to increase the rate of false positives for two reasons. First, we
get a false positive from 
an infrequent line that shares the same counter with a frequent line.
Second, we get false positives from a collection of infrequent lines 
that share the same counter and jointly have too many occurrences.

In our experiments on the first PG DVD, 
we process only files' tops and bottoms, and we use
a 23-bit hash with the 3.4~million distinct lines. 
Assume that hashing distributes lines uniformly and
independently across the counters. Then
the probability
that a randomly selected infrequent line will share a counter with 
one of the $\approx 3000$ frequent
lines is estimated as $\approx 3000 \times 2^{-23} = 3.6\times 10^{-4}$.
These few additional false positives should not be harmful.   

\begin{figure}[tb] \centering
\centering\includegraphics[width=0.6\columnwidth]{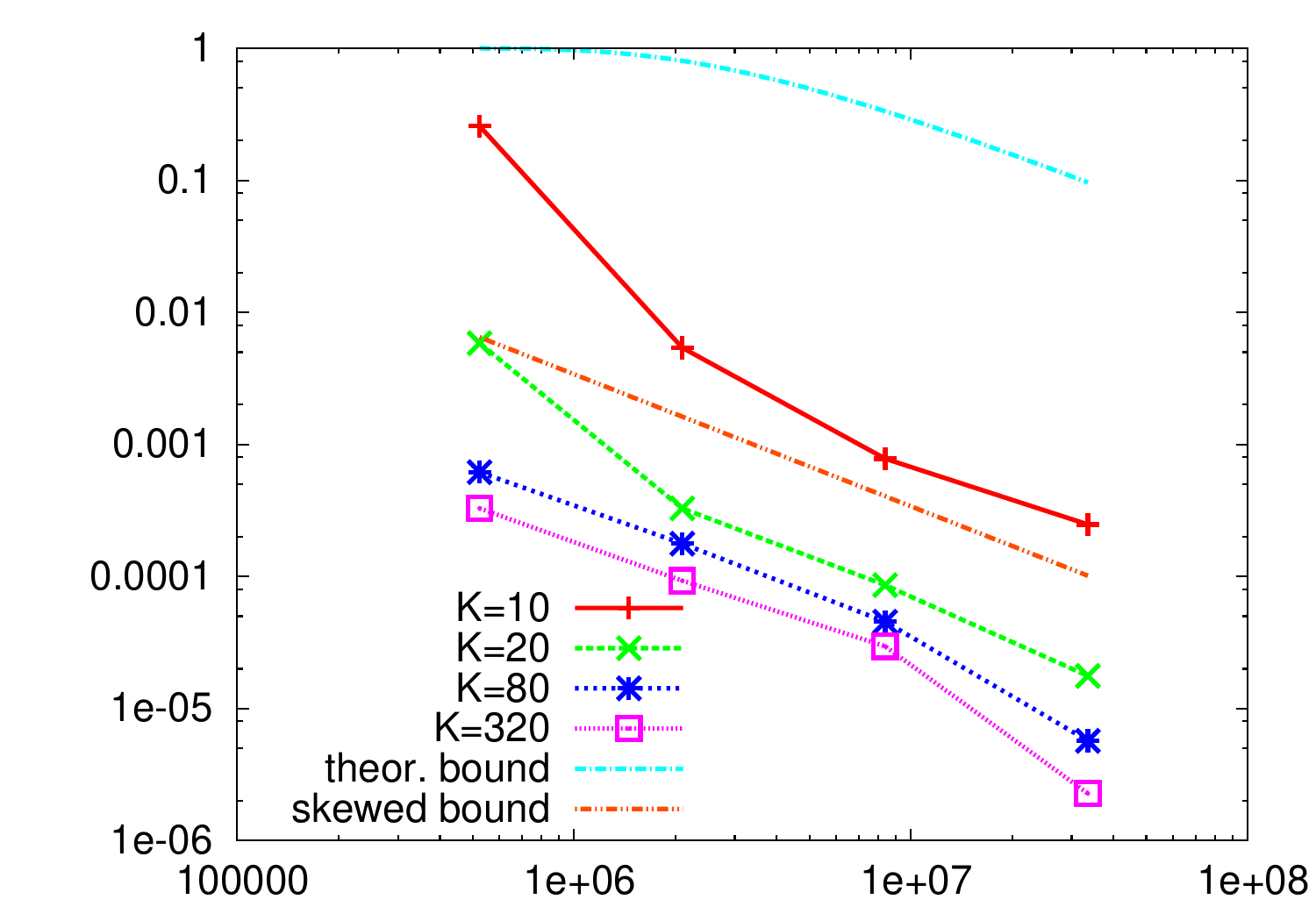}
 \caption{\label{collision-falsepos} Lines falsely assessed as
frequent (as a proportion of the total assessments done) versus
number of counters, $c$.  Several different frequent-line thresholds ($K$)
are shown, as well as the theoretical bound $1-e^{-\frac{n-1}{c}}$
where $n$ is the number of distinct lines. The skewed distribution
bound is $p\frac{n-1}{c} $ with  $p=1/1000$.
}
\end{figure}

It is more difficult to assess the additional false positives arising when
a collection of infrequent lines share a counter and together have 
an aggregate frequency exceeding the frequent-item threshold, $K$. 
We can approach this problem experimentally.
Since we do not precisely know the boilerplate locations for the entire first
DVD, we plot the number of lines that falsely appear frequent due to hashing,
 divided
by the number of lines processed.  This is affected by 
$K$ (we show several values of $K$) 
and by the number
of counters used ($x$-axis).  Fig.~\ref{collision-falsepos} shows 
these experimental results and two theoretical bounds we shall derive.

Consider the probability,
$p_{\mathrm{FP}}$, that a
line $j$ picked at random among all lines
 in a file is a false positive.
Let 
$X_1$ be the frequency of our chosen line. Naturally, more
frequent lines are more likely to be picked.
By definition, if the line is frequent, it cannot be a false
positive.  Thus
$p_{\mathrm{FP}} = P(X_1 \leq K) 
P( \mbox{ line $j$ is a false positive} | X_1 \leq K)$.

Assuming that the hashing is uniform and independent, then the
number $N$ of distinct lines hashed to a certain value follows
a binomial distribution. We need to pick a line $j$ at 
random: for $k\geq 1$, the probability that $k$ 
distinct lines are hashed to the same hash value is 
$P(N=k)= B(n-1,k-1,\frac{1}{c})={{n-1} \choose {k-1}} (\frac{1}{c})^{k-1} (1-\frac{1}{c})^{n-k} $
or, alternatively,
$P(N=k)= \frac{kc}{n}B(n,k,\frac{1}{c})= \frac{kc}{n}{n \choose k} (\frac{1}{c})^{k} (1-\frac{1}{c})^{n-k}$.
Because $n$ is large,
we can approximate the binomial distribution with a Poisson distribution
($e^{-\frac{n-1}{c}} \left (\frac{n-1}{c} \right )^{k-1} /(k-1)!$). 
The probability that we have a false positive when $X_1\leq K$ is
$P( \mbox{ line $j$ is a false positive} | X_1 \leq K)=P(X_1+X_2>K| X_1 \leq K)P(N=2)+P(X_1+X_2+X_3>K| X_1 \leq K)P(N=3)+\ldots$ where $X_i$ denotes
the count of one of the various distinct lines with the same hash
as line $j$. 
Moreover, we can bound how many distinct lines can be hashed together.
Indeed, under the assumption that $n\leq c$,  because
$P(N=k)\approx e^{-\frac{n-1}{c}} \left (\frac{n-1}{c} \right )^{k-1} /(k-1)!
\leq 1/(k-1)!$, we have that  $P(N\geq 9)\lesssim \sum_{k=9}^{\infty} 1/(k-1)! \approx 2\times 10^{-5}$ and $P(N\geq 6) \lesssim \sum_{k=5}^{\infty} 1/(k-1)!\approx 0.05$.

The exact probability of detecting
a false positive naturally depends on the value of the counter $X_1$,
but we can at least bound the probability of observing
a false positive as follows.  
First, a false positive requires one (or more)
lines hashed together with the current line.  Thus, a crude upper bound
is $p_{\mathrm{FP}} \leq P( \mbox{ line $j$ is a false positive} | X_1 \leq K)
\leq P(N \geq 2 ) =
1- P(N=1)
=1-(n-1)!(1-1/c)^{n-1}
\approx 
1-e^{-\frac{n-1}{c}}$.

Finer bounds are possible by
using Chebyshev's inequality or exact knowledge of the distribution
of the $X$ values.
If we assume that the distribution of the $X$
values is very skewed, $X$ is larger than $K$ with a very small probability $p$ and
small compared to $K$ otherwise, 
then we can also assume that $P(\sum_{i=1}^k X_i >K|X_1\leq K) \lesssim (k-1) p$
for $k<10$.
Then, the probability of a false positive when $X_1\leq K$ is 
$P(X_1+X_2>K|X_1\leq K)P(N=2)+P(X_1+X_2+X_3>K|X_1\leq K)P(N=3)+\ldots 
 \lesssim \sum_{k=2}^\infty (k-1) p e^{-\frac{n-1}{c}} \left (\frac{n-1}{c} \right )^{k-1} /(k-1)!
=p \frac{n-1}{c}$.  
(The expected value of the Poisson distribution is $\sum_{k=0}^\infty k e^{-\lambda} \lambda^k /k!= \lambda$.)
Hence, $p_{\mathrm{FP}}\leq p\frac{n-1}{c} $  
and it follows that if  $P(\sum_{i=1}^k X_i >K|X_1\leq K) \lesssim (k-1) p$
for $k < 10$, the probability of a false positive is
no more than $p$. We estimate 
from Fig.~\ref{probahisto} that the 
Gutenberg line data set has such a distribution with $p\approx1/1000$ for $K\approx 100.$
In  Fig.~\ref{collision-falsepos} this bound, $p\frac{n-1}{c}$, is 
compared against some experimental results.

\subsubsection{Tracking Hot Items}

Many algorithms have been
developed for detecting ``frequent items'' in streams, and we might
use one with Algorithm~\ref{boiler-algo}.  In such 
a context, we are not interested in counting how many times a
given item occur, we only want to retrieve frequent items.
Cormode and Muthukrishnan survey some of them~\cite{corm:whats-hot}. 
Given enough memory, Pass~1 of our algorithm can be done with one
pass through the corpus.
Yet if we seek to reduce memory usage, then we must either
give up the ability to capture all frequent items (create false negatives), 
or the ability to capture only the frequent items (create false positives), 
or several passes are required.

Typically, in one pass,
these algorithms deliver a set that contains all frequent items (and some
infrequent items).  
A particularly simple and fast deterministic method,
\emph{Generalized Majority} (GM), has
been developed independently by several
authors~\cite{dema:freq-est-esa02,karp:repeats03,DBLP:journals/scp/MisraG82}.

\begin{algorithm}
 \begin{algorithmic}
\STATE \textbf{Input}: a stream of length $n$ 
\STATE initialize all $c$ counters to zero
\FOR { each element in the stream }
\IF{the element is not currently monitored by a counter and a counter is zero}
\STATE set this counter to monitor the item
\ENDIF
\IF{the element is monitored} 
\STATE increment the counter
\ELSE
\STATE decrement all counters
\ENDIF
\ENDFOR
 \end{algorithmic}
\caption{\label{algo:genmajo}GM 
Algorithm~\protect \cite{dema:freq-est-esa02,karp:repeats03,DBLP:journals/scp/MisraG82}: at the 
end of the run, all elements occurring at least $n/(c+1)$ times are monitored by a counter.}
\end{algorithm}

GM uses $c$ counters, 
where $c \geq 1/f -1$ and $f$ is the minimum (relative) frequency of a frequent
item; in the special case where $f=1/2$, a single counter is sufficient 
(see Algorithm~\ref{algo:genmajo}). 
GM can be implemented to run in $O(1)$ time per element~\cite{dema:freq-est-esa02}.  
The implementation uses doubly linked lists of
doubly linked lists, as well as auxiliary data structures mapping counted items to
their counters. Hence it may use significant memory per counter. 

Probabilistic variants of GM have been proposed~\cite{dema:freq-est-esa02}. However,
in the case where the distribution is very skewed, the algorithm
already provides a good approximation of the frequent items: it suffices
to keep only the items with the largest count values.  We believe this observation is novel, though there is related work~\cite{cormode2005sam}.
To evaluate its accuracy, we generated generalized 
Zipfian data sets ($n\approx $100,000)
containing 1,000 distinct items: the frequency of the rank $r$ item
is proportional to
$1/r^x$. We then apply the GM algorithm and find the largest value of $k$ such that
the top-$k$-items query is exact. 

\begin{figure}\centering
\centering\includegraphics[width=.7\columnwidth]{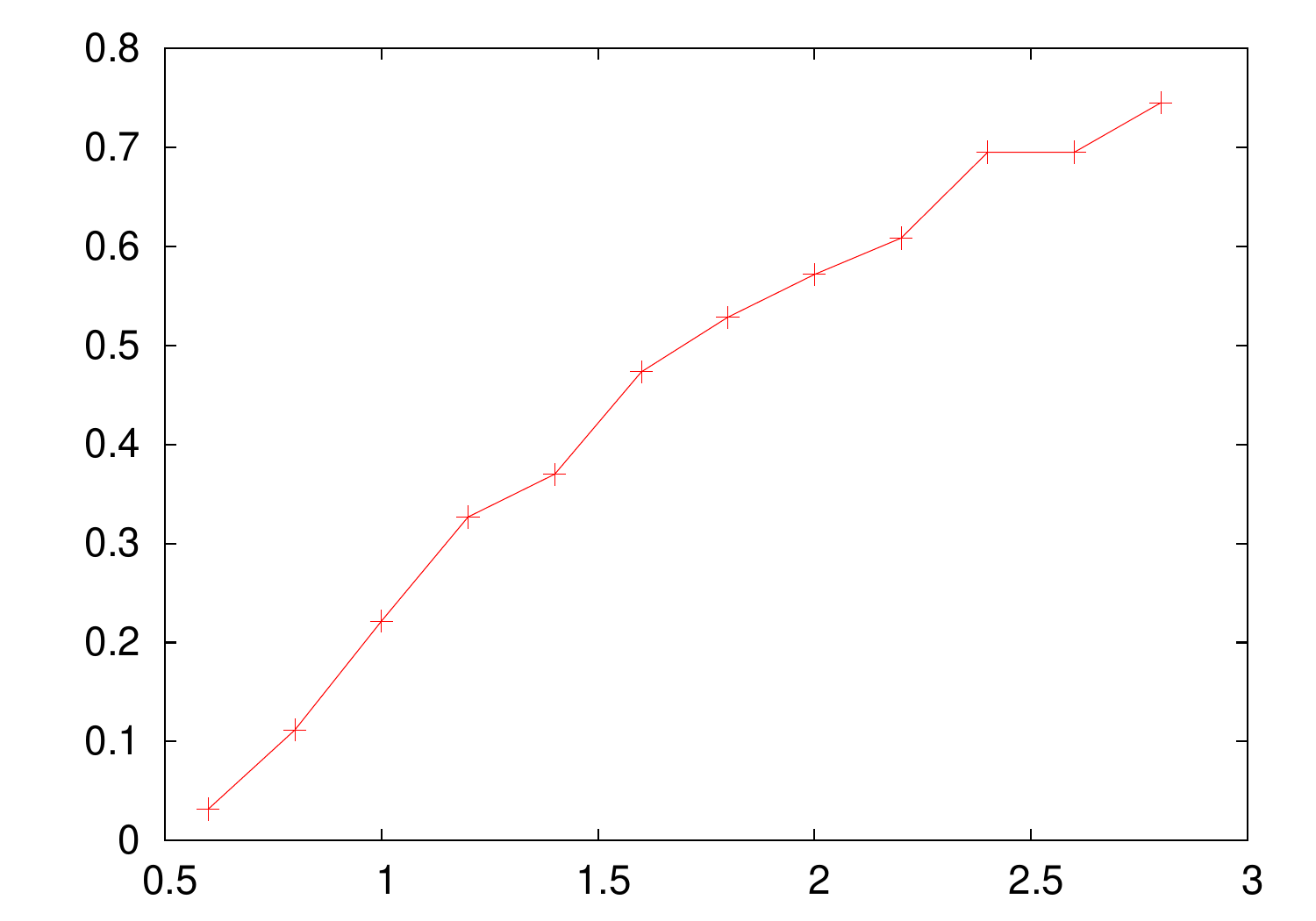}
 \caption{\label{demainefficiency}Efficiency of the GM
algorithm for various Zipfian distributions.  The item of
rank $r$ has frequency proportional to $1/r^x$. }
\end{figure}

Let this value of $k$ divided
by the number of counters $c$ be called the \emph{efficiency} of the algorithm.
For each value of $x$, we repeat the experiment 30 times, with 
different random ordering of the items, and compute the average
efficiency. The result is presented in Fig.~\ref{demainefficiency}
for $c=30$; we determined empirically that
 the efficiency is not sensitive to the number 
of counters. Clearly, for highly skewed  distributions, 
the efficiency can be very close to 100\%.

Consider using GM to detect frequent lines in Algorithm~\ref{boiler-algo}.
Scanning only the tops and bottoms of files, we see  $3\times 10^6$ lines, 
and suppose we choose $K=100$.
To guarantee a set of items that includes all such frequent
lines we need at least $3\times 10^4$ counters 
( $3\times 10^4 \times 100 \geq 3 \times 10^6$).   
Since we know that this is much larger than the number of
frequent items, we will have introduced many false positives.
However, by the previous argument and assuming a very 
skewed distribution, we can select the items with 
a relatively large count and expect to have few false positives.
We will review the applicability of GM experimentally later on.

\subsection{Heuristic Improvements}
\label{heuristic-rules}

A large majority of PG e-books can have their preambles and
epilogues detected by a few heuristic tricks.  However, there are many 
exceptions where the tricks fail, and our experience is that they 
cannot replace the frequent-line approach
without being significantly more complex and constantly updated.
Yet, heuristics can improve processing based on frequent lines.  

The heuristic rules we consider can be expressed as Java
regular expressions.

\begin{itemize}\setlength{\itemsep}{0ex}\setlength{\parskip}{0ex}

\item If a line near the beginning of the file matches the following, it is part of the preamble.\\
{\tt \footnotesize \verb+[\s*]+*\verb+\*[\s]+?(START\verb+\s+OF\verb+\s+(THE|THIS)\verb+\s+PROJECT\\\verb+\s+GUTENBERG|END[\verb+\s+*]THE\verb+\s+SMALL\verb+\s+PRINT!).*}

\item If a line near the end of the file matches the following, it is part of the epilogue.\\
{\tt \footnotesize This|THIS|this|Is|IS|is|The|THE|the|Of|OF\\|[*\verb+\s+]|of)*(End|END|end)(\verb+\s+|Of|OF|of|The|THE\\|the|This|THIS|this)*(Project\verb+\s++Gutenberg|\\PROJECT\verb+\s++GUTENBERG).*}

\item If a line begins with \texttt{ETEXT} and is near the end of the file, it is part of the 
epilogue.

\end{itemize}

\section{Experimental Results}
\label{exper}

We implemented Algorithm~\ref{boiler-algo} and the data 
structures discussed in \S~\ref{data-structures} in Java 1.5
(using Sun's JDK 1.5.0)
and tested them on a older machine  with Pentium~3
Xeon processors (700\,MHz with 2\,MiB cache)
and 2\,GiB of main memory. The OS kernel was 
Linux~2.6.18, and the disk system had three 10\,kRPM Ultra SCSI
disks. 
We copied the contents of the first PG DVD onto the hard disk and
did some simple preparations, primarily unzipping all files and
removing README files.

We chose 100~e-book files randomly out of the approximately 11,000
and the boilerplate boundaries were determined exactly by the authors. 
The algorithm then processed all files ($\approx 11,000$), and then its accuracy
was assessed by checking the sample.
The GM data structure used $c=10^5$ counters, whereas the hashing algorithm
used $c = 2^{23}$ counters. (Since GM stores lines and uses
complicated auxiliary data structures, the actual memory
used by GM and hashing was much closer than comparing $c$ values would
indicate.) 
The threshold beyond which a line is considered frequent is $K=10$
for all data structures except for GM where we used a 
counter value\footnote{GM decrements, so this does not mean $K=5$.}
 of 5.

Results were compared with and without the use of the supplementary
heuristics from \S~\ref{heuristic-rules}.

\subsection{Errors}

\begin{figure*}[htb!]
\centering
\subfigure[\label{hdr-noheur}Preamble errors. ]{
\centering
\includegraphics[width=.45\textwidth,angle=0]{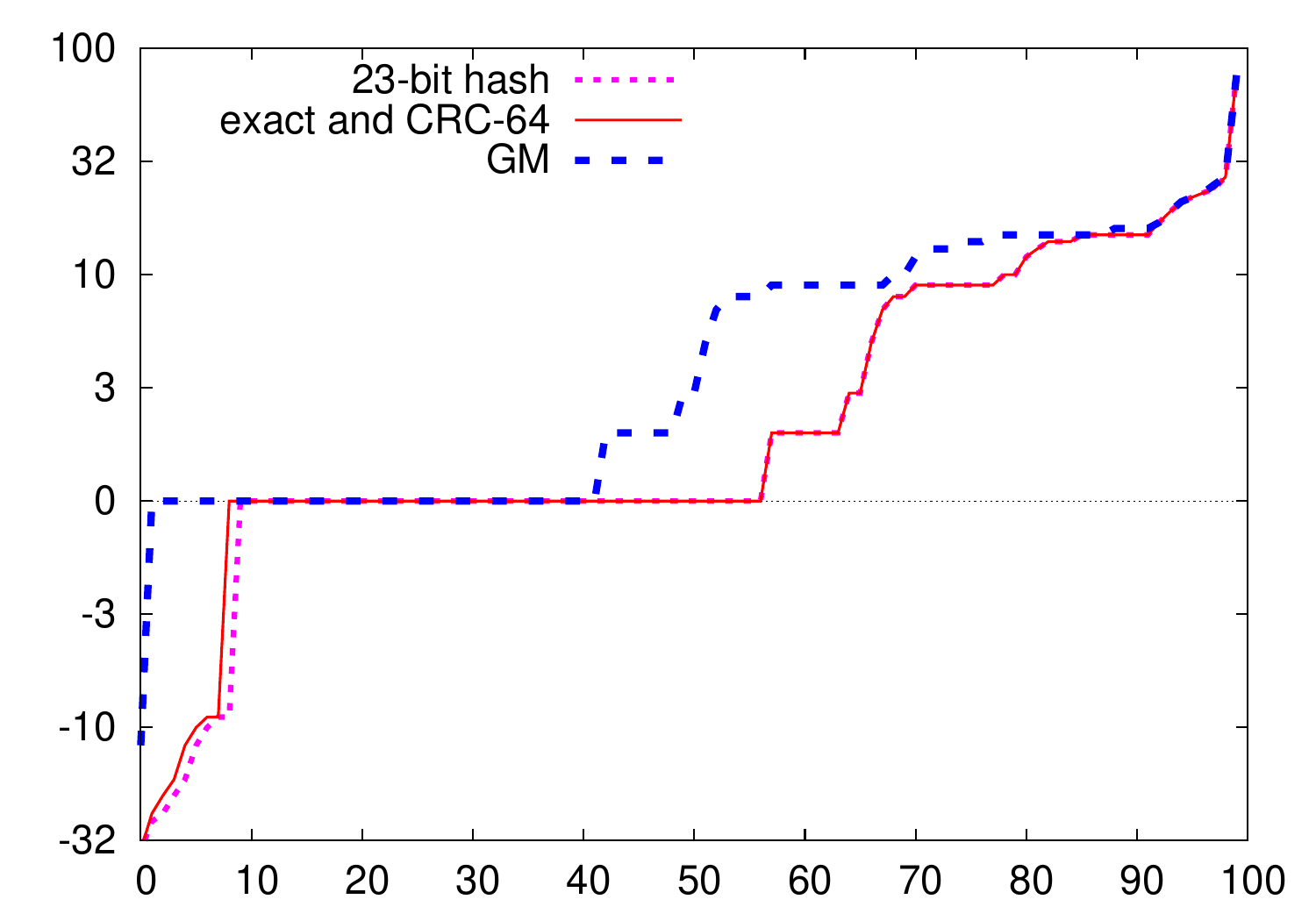}
}
\subfigure[\label{ftr-noheur}Epilogue errors.]{
\includegraphics[width=.45\textwidth,angle=0]{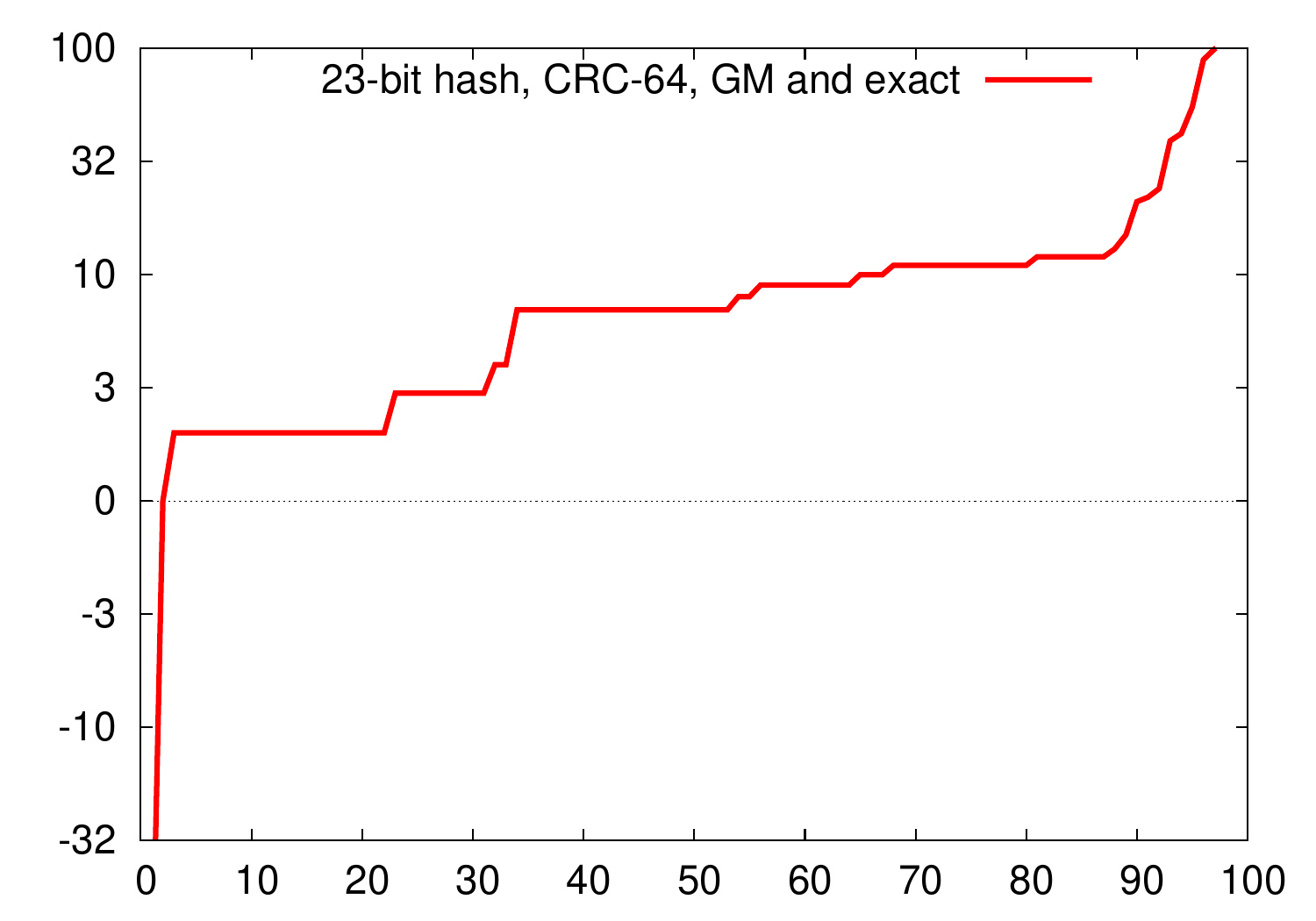}
}\\
\subfigure[\label{hdr-heur}Preamble errors. 23-bit-hashing (not shown) was
marginally better than CRC-64 on the first 10 e-books and identical for the rest.]{
\includegraphics[width=.45\textwidth,angle=0]{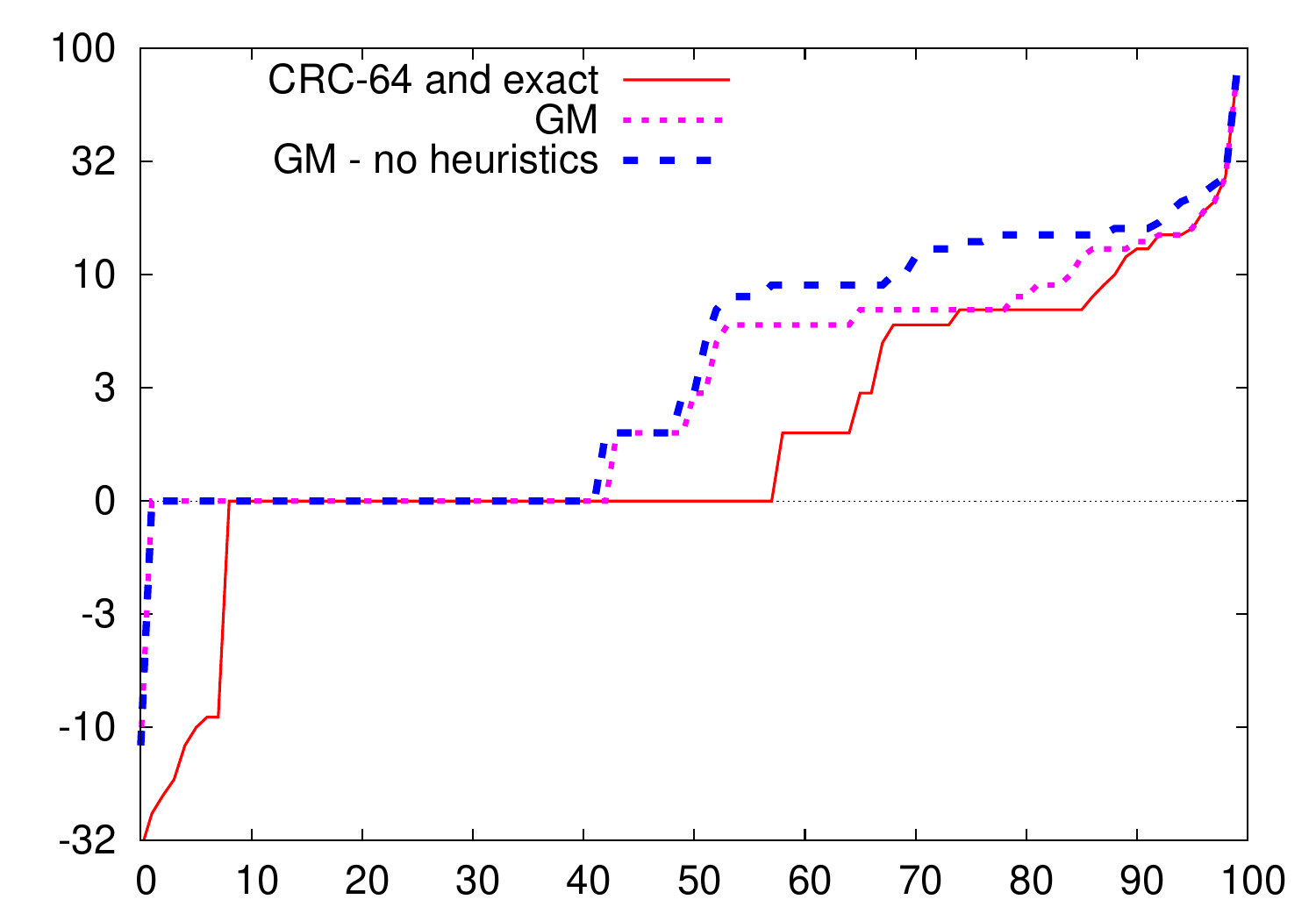}
}
\subfigure[\label{ftr-heur}Epilogue errors.]{
\includegraphics[width=.45\textwidth,angle=0]{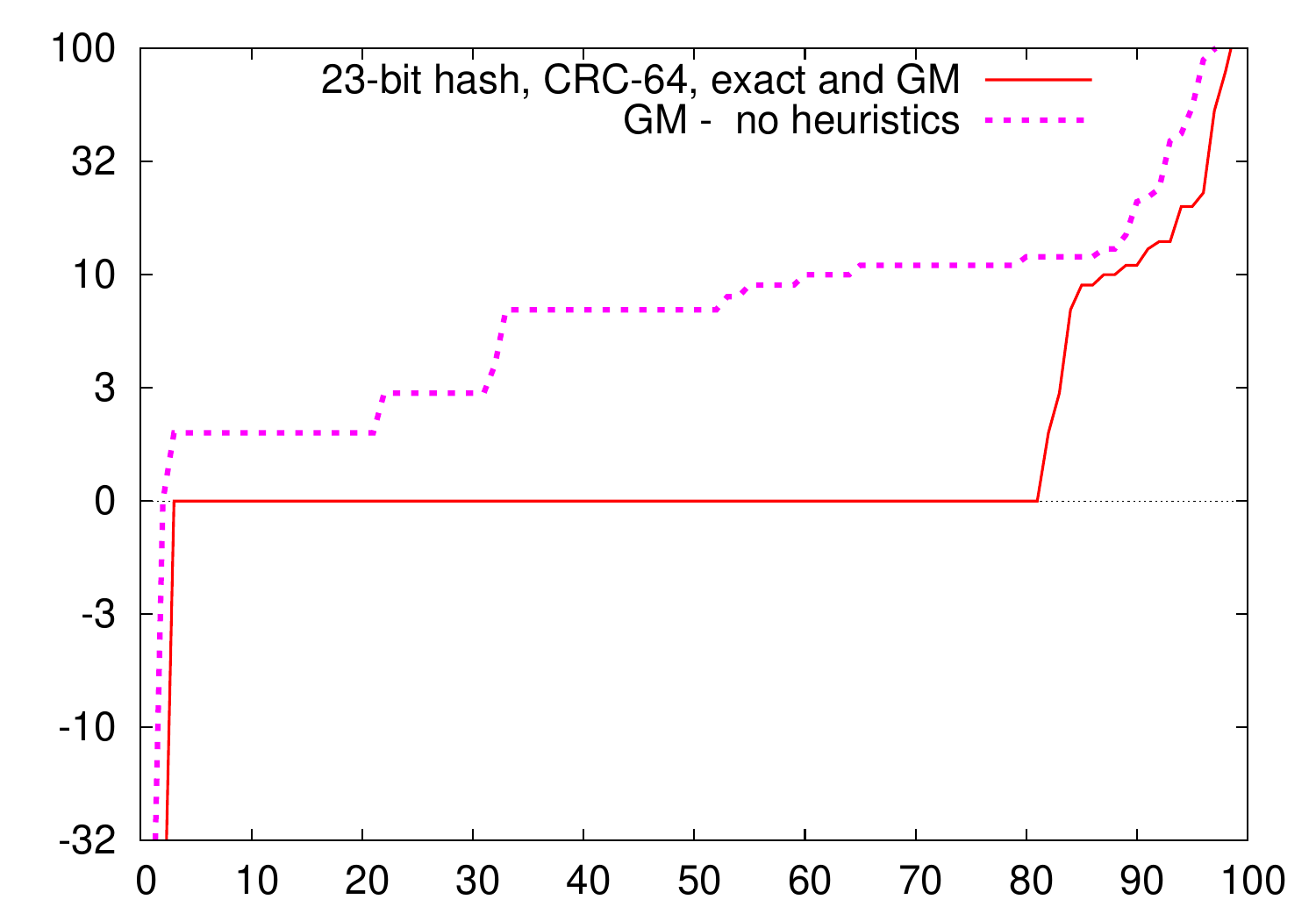}
}
\caption{\label{preepierrors}Preamble/epilogue errors (negative quantities
mean that too many lines were removed; positive quantities mean insufficient
lines were removed). 
In \ref{hdr-heur}~and~\ref{ftr-heur} heuristics were
 used along with the frequent-line algorithm (with the exact and GM
data structures) whereas heuristics were not used in \ref{hdr-noheur} and \ref{ftr-noheur}. 
For comparison, GM is also shown without heuristics.}
\end{figure*}

Figures~\ref{hdr-noheur}~and~\ref{ftr-noheur} show the number of lines
missed by the preamble-removal or epilogue-removal process, respectively,
for each of the 100~files where we sorted the files by the number of lines missed.
Errors (number of lines) are measured
in actual e-book lines, including trivial lines that pre-processing had
discarded. 

In these experiments we did \textbf{not} use the regular-expression-based heuristics.
Looking at epilogues, we see the choice of data structure did not have much
effect on accuracy; problems with false positives or negatives must come
from the data itself.  For preambles, we see that the GM
approach had moderately higher errors in about 30\% of the cases.  However,
this always involved 10 or fewer lines.

Figures~\ref{hdr-heur}~and~\ref{ftr-heur} show how the errors changed
when  we \textbf{did} use the regular-expression-based heuristics.
Comparing results on preamble detection, we see that the heuristics
were somewhat helpful, but GM still had
difficulties compared to using exact counts
in about 30\% of the cases. 

Comparing results on epilogue detection, we see that heuristics reduce
our error to 0 in about 75\% of cases (whereas in these cases,
GM without heuristics would have had an error of 1--10
lines).

\subsection{Run Times}

Our Java implementation of Algorithm~\ref{boiler-algo} 
processed the DVD's files in 20-30 minutes, for each of
the data structures. 
The bottleneck is our unoptimized
implementation of line pre-processing; this trivial task obscures
the differences between the data structures.
Yet, it seems clear that with sufficient implementation effort this
preprocessing bottleneck could be removed.  Then, the speed
differences would arise from the choice of frequent-lines data structure,
and the differences between these
data structures will be most pronounced during the construction of
the structure (the first pass); during the second 
pass, the classification of lines typically involves a similar O(1)
frequency look up per line.

\begin{figure}[htb]
\centering
\includegraphics[width=0.6\columnwidth,angle=0]{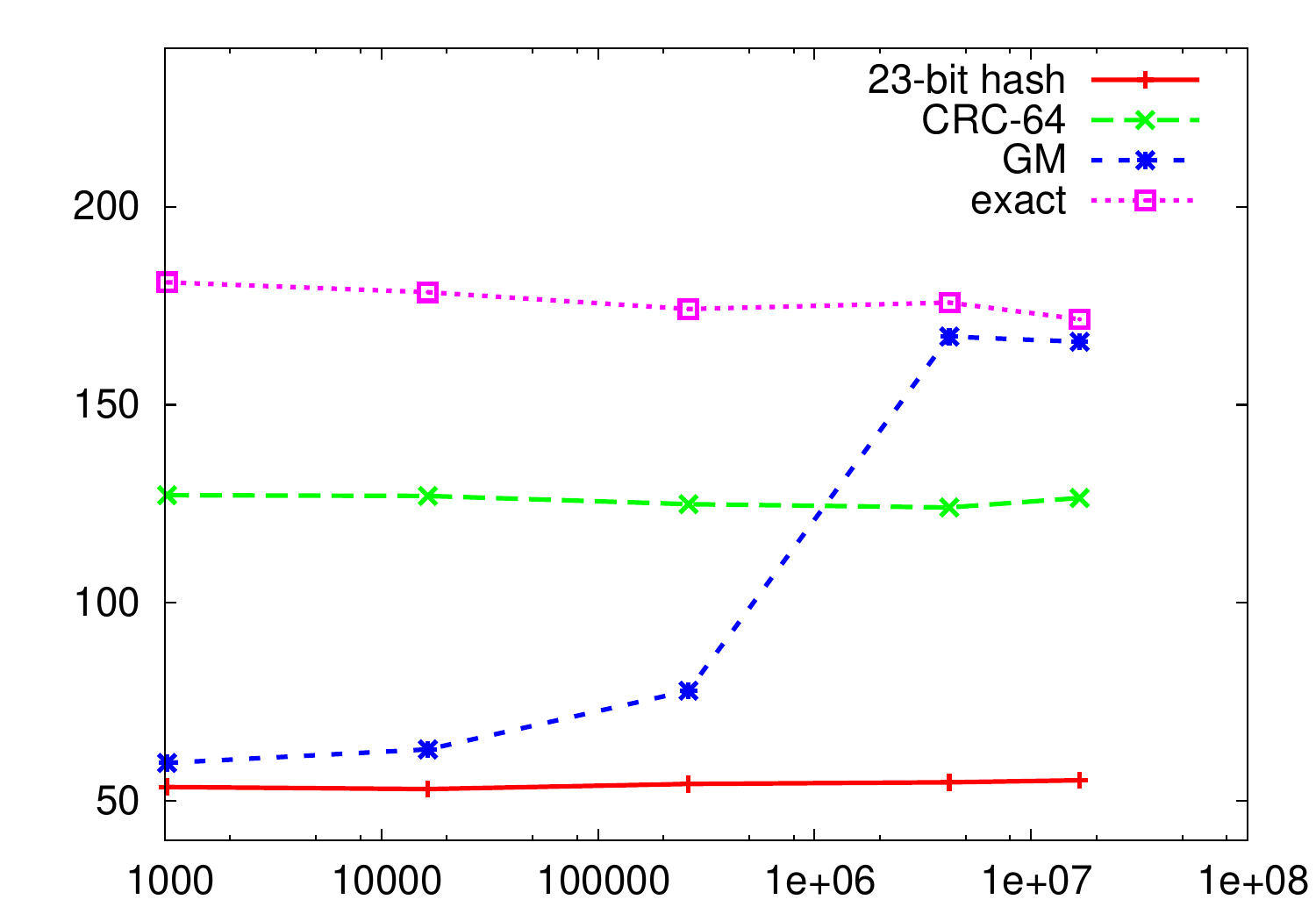}
\caption{\label{timesfig}Wall-clock times (s) to build the
frequent-lines structures vs. the number of counters, $c$.}
\end{figure}

Therefore, we prepared a single file containing pre-processed tops
and bottoms of the files on the first PG DVD.  We then used this
file to construct each of the data structures, subtracting the
time  (approximate 20\,s) it took to read the file.   
Experiments considered a range of values for $c$, the number
of counters used.
For each data point, ten trials were made and their average is
shown in Fig.~\ref{timesfig}.
Exact and CRC-64 always use as many counters as there
are distinct lines whereas the 23-bit hash always allocates the same
number of counters; 
times for these data structures should not
vary with $c$.

We see that 23-bit hashing was the fastest approach,
requiring about 55~s.  For small $c$, GM is nearly as
fast, but it slowed as $c$ increases to the number of
distinct items (3.4~million), at which point its speed resembles the 
exact structure.  This is not surprising; GM uses a Java \texttt{HashMap}
to associate each counted line to its counter.  The ``exact''
implementation was slowest, and it 
also uses a \texttt{HashMap} for the same purpose, although
``exact'' maps to an \texttt{Integer} rather than GM's more complicated
data structure.  When fewer counters are used, GM maintains a smaller
\texttt{HashMap} and, despite having to do additional book-keeping beyond
``exact'', the speed benefit of the smaller \texttt{HashMap} outweighs
the cost of book-keeping.   When CRC-64 is used, these values are also
stored in a \texttt{HashMap}, so the speed advantage enjoyed by CRC-64
presumably arises from its more economical use of memory.
If GM stored CRC values rather than strings, its performance 
should match CRC-64 when $c \geq $3.4~million.

Experiments were limited to $2^{24}$ counters because GM made  our system
run out of memory when $2^{26}$ counters were used 
(we limited the JVM to 1.8~GiB).
While this might be due to an inefficient implementation, we feel that only our
23-bit hashing implementation (which uses space-efficient arrays of basic types) 
would successfully scale for much larger corpora.  Of course, using
external-memory techniques is also a viable solution. Recall that 
GNU/Linux shell utilities, 
presumably highly optimized,  could sort 
and build the list of frequent lines in under 100\,s.

\subsection{Comparison to GutenMark}

Of those software tools that 
reformat PG e-books,
it appears
only GutenMark~\cite{GutenMark} formally attempts to detect the preamble,  so it 
can be stripped.  We used its most recent production release, dated 2002,
when PG e-books did not have a long epilogue.  Thus we can only test
it on preamble detection.

GutenMark  attempts to bypass all prefatory material, whether inserted
by PG or part of the original text.  To do so, it tries to detect
the body of the text, looking for markers such as  ``Chapter I.''
Unlike our tool, it thus tries to omit tables of contents, etc.

With GutenMark,  the preamble is stripped only in an output format
(we use HTML) and the stripping is accompanied with various reformatting 
heuristics: all-capitalized text is transformed in italics, punctuation is transformed
into HTML entities, and headers
are identified and marked using h1 elements in HTML.

To determine where GutenMark infers the preamble, we use the main header
tags (\verb+h1+ elements in HTML) that it generates.  Usually, the first
header is ``Prefatory Materials'' and the second begins the body of the
e-book.   We then have to map this back to corresponding lines in the
original text file, which may look quite different due to reformatting.

Processing e-book files with GutenMark is slow 
compared to  our implementations of
Algorithm~\ref{boiler-algo}, but it is unfair to draw
conclusions from this.  GutenMark was not intended to
process a large corpus, and  it processes all of the file.
 A heuristic-based approach similar to GutenMark
could probably be implemented so that it would run
faster than our approach: it has no frequent-line
data structure to maintain and can probably process
the corpus in a single pass.  However, is it accurate?

\begin{figure}[htb]
\centering
\includegraphics[width=0.6\columnwidth,angle=0]{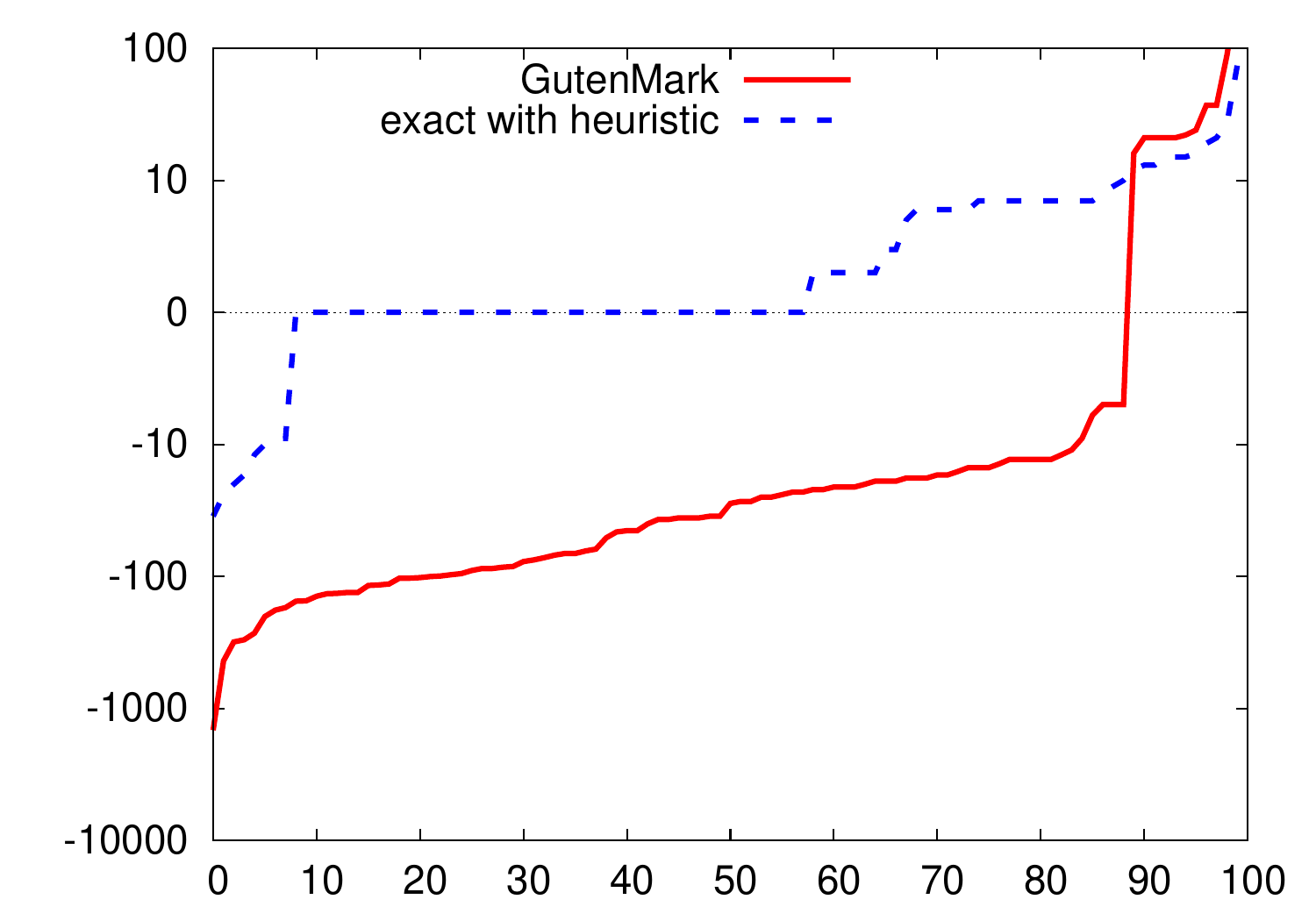}
\caption{\label{gutenmark-errs} Preamble  errors for GutenMark.}
 \end{figure}

Figure~\ref{gutenmark-errs} shows the errors obtained when we inferred
where GutenMark detected preambles.  In one case, \textit{Diary of Samuel
Pepys, October 1666}, we see an error of more than 1000 lines.  
Apparently the diary format used did not have headings GutenMark
could detect.

Despite several other large errors, in many cases
the GutenMark approach worked reasonably well.  
Its  erratic behaviour illustrates the risk of
heuristics rules inferred manually from an out-of-date
corpus version.

\section{Conclusion}

%
%

The PG Project is a good example of a large, poorly structured data set.
Nevertheless, it is evidently proving useful since thousands of volunteers
contribute to the project every year.

Detecting the PG-specific preambles and epilogues is maybe
surprisingly difficult. There are instances where a human
without knowledge of English probably could not accurately determine
where the preamble ends.  Nevertheless, our approach based on line
frequency can approximately
(within 10\%) detect the boilerplate in more than 90\% of the
documents (see Fig.~\ref{preepierrors}).

Line frequency follows a very skewed distribution and thus, as we have
demonstrated, hashing to small number of bits will not lead to  a large
number of lines falsely reported as frequent.
Indeed, using 23-bit line hashing, we can approximately find the frequent
lines, with an accuracy sufficient so that preamble/epilogue detection is
not noticeably affected. 
Simple rule-based heuristic can improve accuracy in some cases, 
as observed with epilogues. 

Future work should include automatic detection of the remaining cases,
which require human intervention.  In general, this will
be difficult.  Yet a few simple sanity tests
are possible; for instance, seeing `Gutenberg' in 
the body of an e-book is typically  a sign of trouble.

All that is required for our approach to scale up
is to detect frequent lines efficiently. Because
the distribution of line frequency is very skewed,
highly efficient algorithms are possible, such as
the hashing and GM algorithms, without sacrificing
accuracy significantly.

Training accuracy could be improved by including only one copy of each
e-book, or forming separate frequent-lines data structures
for preambles and epilogues.
Much replication can be removed by using PG catalogue metadata
that is available from the Project.  We did not use it; thus, our results
apply to similar corpora lacking catalogue metadata.

\section*{Acknowledgements}
The first author was supported in part by NSERC grant 155967, and the
second author was supported in part by NSERC grant 261437 and FQRNT grant 112381.

\balance
\bibliographystyle{alpha}
\bibliography{lemur}

\end{document}